%% file: rfi.tex
\newcommand{\TSM}{T_{\rm{SM}}}

\newcommand{\mdm}{m}

\def\rh{\rho_{\rm dm}}
\def\nh{n_{\rm dm}}
\def\mh{m}
\def\rsm{\rho_{\rm sm}}
\def\tsm{T}
\def\ssm{s_{\rm sm}}
\def\mpl{M_{\rm pl}}

\documentclass[reprint,amsmath,amssymb,aps,prd,]{revtex4-1}

\usepackage{graphicx}
\usepackage{dcolumn}
\usepackage{bm}
\usepackage{hyperref}
\usepackage{comment}

\usepackage[hyperref,svgnames]{xcolor}

\newcommand{\gsim}{\lower.7ex\hbox{$\;\stackrel{\textstyle>}{\sim}\;$}}
\newcommand{\lsim}{\lower.7ex\hbox{$\;\stackrel{\textstyle<}{\sim}\;$}}

\newcommand{\kpc}{{\, {\rm kpc}}}

\newcommand{\eV}{{\, {\rm eV}}}
\newcommand{\keV}{{\, {\rm keV}}}
\newcommand{\MeV}{{\, {\rm MeV}}}
\newcommand{\GeV}{{\, {\rm GeV}}}
\newcommand{\TeV}{{\, {\rm TeV}}}

\def\beq{\begin{equation}}
\def\eeq{\end{equation}}
\def\bea{\begin{eqnarray}}
\def\eea{\end{eqnarray}}
\def\bitem{\begin{itemize}}
\def\eitem{\end{itemize}}
\newcommand{\bec}{\begin{center}}
\newcommand{\eec}{\end{center}}
\newcommand{\ba}{\begin{array}}
\newcommand{\ea}{\end{array}}
\def\inv{^{\raise.15ex\hbox{${\scriptscriptstyle -}$}\kern-.05em 1}}
\def\lbar{{\lower.35ex\hbox{$\mathchar'26$}\mkern-10mu\lambda}} %lambda bar

\def\to{\rightarrow}

\let\<=\langle
\let\>=\rangle

\let\+=\uparrow

\let\al=\alpha

\newcommand{\td}{{\widetilde T}}
\def\Teq{T_{\rm EQ}}

\def\tdec{t_{\rm dec}}
\def\TDM{\widetilde{T}}
\def\lc{l_c}

\def\lFS{l_{\rm FS}}
\def\gst{g_{*}}
\def\sv{\langle \sigma v \rangle}
\def\Neff{N_{\rm eff}}
\def\DNeff{\Delta \Neff}

\def\GeV{\text{ GeV}}

\def\ltot{l_{\rm tot}}

\def\beq{\begin{equation}}
\def\eeq{\end{equation}}

\begin{document}

\title{Reproductive Freeze-In of Self-Interacting Dark Matter }

\author{John March-Russell}
 \email{John.March-Russell@physics.ox.ac.uk}

\author{Hannah Tillim}
\email{hannah.tillim@physics.ox.ac.uk}
\affiliation{Rudolf Peierls Centre for Theoretical Physics, University of Oxford, Beecroft Building, Oxford OX1
3PU, United Kingdom}

\author{Stephen M. West}
\email{stephen.west@rhul.ac.uk}
 \affiliation{Royal Holloway, University of London, Egham, Surrey, TW20 0EX, United Kingdom}

\date{\today}

\begin{abstract}We present a mechanism for dark matter (DM) production involving a self-interacting sector that at early times is ultra-relativistic but far-underpopulated relative to thermal equilibrium (such initial conditions often arise, e.g.,  from inflaton decay).  Although elastic scatterings can establish kinetic equilibrium we show that for a broad variety of self-interactions full equilibrium is never established despite the DM yield significantly evolving due to $2\to k$ ($k>2$) processes (the DM carries no conserved quantum number nor asymmetry).   During the active phase of the process, the DM to Standard Model temperature ratio falls rapidly, with DM kinetic energy being converted to DM mass, the inverse of the recently-discussed `cannibal DM mechanism'. As this evolution is an approach from an out-of-equilibrium to equilibrium state, entropy is not conserved. Potential observables and applications include self-interacting DM signatures in galaxies and clusters, dark acoustic oscillations,  the alteration of free-streaming constraints, and possible easing of $\sigma_8$ and Hubble tensions.

{\scriptsize DOI: 10.1103/PhysRevD.102.083018}

\end{abstract}

\maketitle

\section{\label{sec1}Introduction}

One of the few facts known with certainty about the nature of dark matter (DM) is that its non-gravitational interactions with the Standard Model (SM) sector are, at most, tiny. Though not necessitated by present observations it is entirely possible that there exist non-gravitational DM-sector \emph{self}-interactions,  in the presence of which new phenomena can occur which may relieve a number of astrophysical or cosmological tensions \cite{Spergel:1999mh,Markevitch_2004,Vogelsberger:2012ku,Rocha:2012jg,Peter:2012jh,Zavala:2012us,Tulin_2013,Kahlhoefer:2013dca}.   Of particular interest to us is the fact that such self-interactions inevitably lead to both elastic \emph{and} inelastic, number-changing scattering processes.  We here study a new DM relic density generation mechanism involving such inelastic self-interactions.

Specifically, we show that if the initial state of the DM sector is ultra-relativistic, $\langle E_{\rm dm} \rangle \gg \mh$, but far-underpopulated relative to full thermal equilibrium, a calculable non-linear process, that we refer to as \emph{Reproductive Freeze-In} (RFI) \footnote{Informal talk by JMR at Galileo Galilei Institute Florence, Summer 2019}, can lead to a DM relic density compatible with current observations, bearing new features of phenomenological import.  During the active phase of the process, the DM yield significantly evolves due to $2\to k$ ($k>2$) processes (for DM without conserved quantum numbers or asymmetry) and the DM sector temperature falls relative to that of the SM bath as DM kinetic energy is converted to DM mass.  Thus RFI is effectively the inverse of the recently-discussed `cannibal DM mechanism' \cite{Carlson:1992fn,Hochberg:2014dra,Hochberg_2015,Pappadopulo:2016pkp,Farina_2016}.

The evolution of a self-interacting DM sector initially possessing a non-zero chemical potential has been studied by a number of groups,
\cite{Heeba:2018wtf,Arcadi:2019oxh,DeRomeri:2020wng,Bernal_2016, mondino2020dark, Evans_2020} with a focus on first finding the conditions under which the initially underpopulated sector fully thermalises and subsequently how this impacts the DM abundance as it freezes-out. In this work, we instead focus on the freeze-in case \cite{Hall:2009bx,McDonald:1993ex,Hall:2010jx} where the DM sector evolution never reaches full equilibrium. We emphasize the simple but crucial fact that, in contrast with an assumption made widely in the literature, entropy per co-moving volume is \emph{not} conserved during the evolution.  This is due to the initial state being a far-from-equilibrium state, the DM relic density in our case being determined by a failure to ever achieve equilibrium.  
To our knowledge, this point has not been previously appreciated.

Note that the assumed far-underpopulated, ultra-relativistic DM-sector initial conditions often result from dynamics of the extreme early Universe. Examples include post-inflation inflaton decay and reheating \cite{Kofman:1996mv,Chung:1998ua,Kolda:1998kc};  the decay of a population of primordial black holes \cite{Argyres:1998qn,Lennon:2017tqq,Hooper:2019gtx} and the decay of superheavy particles associated with, for instance, neutrino mass generation or Peccei-Quinn symmetry breaking. 

As we will discuss, possible signatures of the RFI mechanism include the effects of the associated elastic DM self-interactions in galaxies and galaxy clusters, the signs of dark acoustic oscillations in structure formation, and the weakening of free-streaming constraints on models that produce an initial ultra-relativistic population of DM particles.  Since the RFI mechanism does not require that the DM possesses any exact continuous or discrete stabilising symmetry - just that interactions with the SM are feeble - late-time processes including DM decays occurring in the current universe are a possibility.

\section{\label{sec2}RFI Setup}

A starting assumption of RFI is that the dynamics of the very early universe - post-inflationary reheating, the creation of the universe itself, or some other possible scenario such as the decay of a population of superheavy states, populates both the SM and DM sectors with initial energy densities with ratio $r\equiv \rh / \rsm$. 

In the SM sector, for all particles excepting possibly right-handed neutrinos, the gauge and other interactions are not small, and both kinetic and full species (`chemical') equilibrium are assumed to be achieved rapidly with a `reheat' temperature $T_{0}$.   On the other hand, in the DM sector we assume that the appropriate dimensionless measures of self-interaction strength (which may be a gauge coupling,
a quartic scalar self-interaction, or a combination $(E/M)^n$ involving the typical energy $E$ and the scale $M$ of a higher-dimensional interaction) are such
that elastic scattering is fast compared to the Hubble time, $1/H(t)$.  These interactions quickly establish \emph{kinetic} equilibrium (this simplifying assumption
can be relaxed, see Section \ref{sec:conclusions}) with an associated
kinetic temperature, $\td(t)$, but, in contrast, \emph{number changing} $2\rightarrow k$ $(k>2)$ processes, which drive the system towards \emph{chemical} equilibrium, are parametrically slower.
This is easy to achieve.  For example, if the DM sector consists of a single massive Majorana fermion, $\Psi$, then the
leading interactions are of the schematic form $\Psi^4/M^2 + \Psi^6/M^5+\cdots$, and the ratio of cross sections for the $2\rightarrow 4$ to $2\rightarrow 2$ processes is
$\sim (E /2 \pi M)^4 \times ({\rm logs}) \ll 1$ in the energy regime of interest,  $\mh \ll E \ll M$.
Alternatively, if we consider a single-species scalar model with a perturbative quartic self-coupling $\lambda\lsim 1$, then the ratio of the $2\rightarrow 4$ to $2\rightarrow 2$ processes is $\sim (\lambda^2 /16 \pi^4)\times ({\rm logs})  \ll 1$.   

The distribution functions of particles in kinetic but not chemical equilibrium are given by
\bea\label{eq:dist}
f(p,t) = \frac{Z(t)}{\exp[p/\td(t)] \mp Z(t)},
\eea
where $Z(t) \equiv e^{-\mu/\TDM}$ with $\mu$ the chemical potential associated with deviation from chemical equilibrium, and $\mp$ apply to the Bose/Fermi DM cases. 

A second fundamental assumption of RFI is that \emph{the DM number density, $n_{\rm dm}$ is significantly under-occupied compared to equilibrium during the entire evolution} (we will verify that this assumption is self-consistently correct), corresponding to $Z \ll 1$. If, further, the DM is ultra-relativistic, the number and energy densities in the DM sector are well approximated by
\bea\label{eq:rel_n_and_rho}
n_{\rm dm}(t) &= & \frac{g}{\pi^2} Z(t)\td(t)^3 \label{eq:rel_n},\\
\rho_{\rm dm}(t) &= &  \frac{3 g}{\pi^2} Z(t)\td(t)^4 \label{eq:rel_rho}.
\eea
We will show that although the DM number density can significantly evolve, in an expanding universe full equilibrium may never be reached and a non-trivial DM yield can result.

Although self-interactions within each sector are vital, the RFI mechanism does not utilise SM-to-DM-sector interactions in any essential way, so we assume for pedagogical simplicity that any such interactions are strictly absent.  The SM and DM sectors are then \emph{secluded} from each other and have separate dynamics apart from their gravitational effects upon each other. (For previous work on this see, e.g., \cite{Adshead:2016xxj}.)  Thus during a period where both sectors remain relativistic the energy density ratio, $r$, remains constant up to changes in the number of SM relativistic degrees of freedom, $\gst$, upon
going through a mass threshold for a SM state which we neglect here. In Section~\ref{sec:conclusions} we will comment on these approximations. 

In the DM sector the number-changing processes act to drive $Z(t)$ towards unity, that is, towards chemical equilibrium. Concurrently $\td$ will be falling faster than dictated by Hubble expansion as kinetic energy is converted into mass energy.  Whether the system reaches chemical equilibrium or not depends on the rate of the $2\rightarrow k$ process and how they depend on temperature.  We define the ratio of DM to SM-sector temperatures
\bea\label{eq:Theta}
\Theta(t)\equiv \frac{\td(t)}{\tsm(t)}~,
\eea
which is restricted to lie in the range $\Theta_0\geq \Theta\geq \Theta_{\rm eq}$, where $\Theta_0$ is the initial ratio and $\Theta_{\rm eq}$ is the ratio when the DM sector reaches chemical equilibrium.   As the change in $\Theta$ occurs via processes that are increasing the number of DM states with mass, $\mh$, as soon as the DM becomes non-relativistic, with $\td\simeq \mh$ the development of $\Theta(t)$ ceases for kinetic reasons apart from trivial evolution due to possible changes in $g_*$.  Here we take the DM sector to be comprised of a single type of particle with mass $\mh$.  The generalisation to more complicated DM sectors is straightforward.  As we discuss in the following, we are here for simplicity also assuming that the cannibal mechanism \cite{Carlson:1992fn,Hochberg:2014dra,Hochberg_2015,Pappadopulo:2016pkp,Farina_2016} is frozen out for $\td< \mh$, so the DM density
is determined by RFI.  It is again straightforward to relax this assumption.
Constancy of $r$ in the relativistic regime implies
\bea\label{eq:zvszeta}
\frac{Z(t) \Theta(t)^4}{g_*(t)} =\frac{\pi^4 }{90 g} r~.
\eea

The ratio of DM-sector to SM-sector entropy densities is not constant during the evolution, as the DM sector starts in a far-from-equilibrium state, and the production of DM particles by the inelastic $2\rightarrow k$ processes increases the DM entropy, with the result that $s_{\rm dm}/s_{\rm sm} \propto 1/\Theta(t)$. Instead, it is the ratio of energy densities $r$ that is approximately constant (up to $\gst$ thresholds as mentioned previously). 

The equilibrium value of $\Theta$ is 
\bea\label{eq:Thetaeq}
\Theta_{\rm eq} & \simeq &\left(\frac{\pi^4 r}{90} \frac{g_*}{g} \right)^{1/4} \\
&\simeq & 0.3\;\left(\frac{r}{10^{-4}}\right)^{1/4}\left(\frac{g_* /g}{10^{2}} \right)^{1/4}~;
\eea
though, in all cases of interest to us, evolution of $\Theta(t)$ will stop well before this value is reached.
It is also instructive to estimate the maximum possible size of $\Theta_0$ in order understand the conceivable range over which $\Theta$ can vary.  
As the initial average energy per particle in the DM sector is $\langle E \rangle = \rh^{0}/\nh^0$, one finds
$\Theta_0 \lsim \langle E \rangle/3\tsm_0$
where $\tsm_0$ is the initial temperature of the fully-thermalised SM sector.  Then the most extreme case
with $\langle E \rangle \sim \mpl$ and $\tsm_0\sim 3$\;MeV would have $\Theta^{\rm max}_0\sim 10^{21}$. 

Assuming the DM sector never reaches full chemical equilibrium, and utilising the leading expression for the number density, $\nh$, Eq.~\ref{eq:rel_n_and_rho},
as well as Eq.~\ref{eq:zvszeta} and the value of the SM entropy density, $\ssm= 2\pi^2 g_* T^3/45$, we find the DM yield $Y\equiv\nh/\ssm$ is
related to $\Theta$ by
\bea\label{eq:yieldtheta}
Y = \frac{r}{4\Theta} ~.
\eea
Then, if $\Theta_f$ is the value of $\Theta$ at which the number-changing processes terminate,
the correct DM relic abundance,  $\Omega_{\rm dm} h^2 \simeq 0.119$, implies
$\mh Y_f \simeq 0.434~{\rm eV}$ and thus
\bea\label{eq:Thetafinal}
\Theta_{f}\simeq 57 \left(\frac{r}{10^{-4}}\right)\left( \frac{\mh}{\MeV} \right)~.
\eea

Using $T=\td/\Theta$ and Eq.~\ref{eq:yieldtheta}, one finds that for the correct DM density, the SM temperature at the end of DM particle production must satisfy
\bea\label{eq:TSMfinal}
T_f\geq \frac{4\mh Y_f}{r} \simeq \frac{1.74}{r} \eV ~,
\eea
with the inequality being saturated if the yield changes fast compared to $H(T)$ when $\td \sim \mh$.  We can immediately find a simple upper limit
on $r$ by calculating the contribution of the DM sector to $\Neff$ at the time of Big Bang Nucleosynthesis:
\beq
\DNeff = \frac{4}{7}\gst r~;
\eeq
if we require that this not exceed the conservative limit of $0.3$ \cite{PhysRevD.98.030001} the bound on $r$ is
\beq\label{eq:constr:DNeff}
r < 0.05~,
\eeq
translating into a lower limit $T_f \gtrsim 35 \eV$.  Typically, however, there are other, stronger constraints on $r$.
For example, since there exist strong constraints on either the DM free-streaming scale (if the DM is collision-less) or the DM sound horizon (if the DM is elastically scattering and behaving as an adiabatically expanding gas) arising from structure formation, the physics of an initially ultra-relativistic DM sector is constrained.  We address this constraint on $r$, and thus $T_f$, in Section~\ref{sec:pheno}.

However this analysis hides the fact that Eq.~\ref{eq:TSMfinal} is only true \emph{if} a suitable $Y_f$ giving the observed DM density is achieved at
the conclusion of the mechanism.  This requires an analysis of the Boltzmann equation determining the evolution of $Y$ (and $\Theta$ via Eq.~\ref{eq:yieldtheta}).

\section{\label{sec3}The RFI  Yield Equations}

We are interested in the increase in the DM number density in the presence of number-changing interactions.  We focus on the single species case for pedagogical simplicity (the more general case of multiple DM particle species of different masses interacting in number-changing ways is a direct, if messy, generalisation).   Then, the evolution of the single relevant phase space distribution function, $f(E,t)$, is governed by the Boltzmann equation, which, once integrated, gives the time derivative of the particle number density\cite{Kolb:206230}:
\begin{equation}\label{eqn:Boltzmann_n}
\dot{n}(t) + 3 H n = \frac{g}{(2\pi)^3} \int \mathrm{\bf C}[f] \frac{d^3 p}{E} \equiv c_{n} ~.
\end{equation}
The so-called collision term on the right-hand side sums over all number-changing $2\rightarrow k$ and $k\rightarrow 2$ interactions involving the particle in question.  Since by assumption the elastic interactions are `fast' and maintain kinetic equilibrium, we do not need to explicitly include the effect of $2\rightarrow 2$ processes once the semi-thermal form of the distribution functions, Eq.~\ref{eq:dist} is used.  
Thus each of the terms is an integral, over external momenta, of the number-changing matrix element and a combination of the phase space distributions of the external particles.   In the ultra-relativistic regime $\td\gg \mh$ of interest, the dependence of the collision term on the DM sector temperature, $\td$, and the energy-independent under-occupancy factor, $Z$, can be written as 
\begin{equation}\label{eq:collision}
c_n =  g \td^4 \sum_{k>2}   ~(k-2) Z^2 (1-Z^{(k-2)})~{\mathcal I}_k( \td/M )~,
\end{equation}
where the sum on $k$ runs over all number-changing channels, and the functions ${\mathcal I}_k(\td/M)$ represent the strength and non-trivial temperature dependence of the $2\rightarrow k$ interaction. Here $M$ is a possible energy scale, such as that defining an interaction strength in the effective theory, and the presence of the $Z^{(k-2)}$ term follows from considerations of detailed balance. 

The first simplification of Eq.~\ref{eq:collision} results from the fact that over the whole range of non-trivial
evolution of $n(t)$ the DM state is far under-occupied, with $Z\ll 1$ so we may drop the inverse $k\rightarrow 2$ process term $Z^{(k-2)}$.  Similarly,
irrelevantly small $j\leftrightarrow k$ terms with both $j,k>2$ were already neglected in Eq.~\ref{eq:collision}.  Secondly,  we assume that $\td/M \ll 1$ for all $\td$ of interest  so that we may expand the function ${\mathcal I}_k(\td/M)$ and keep only the leading term.  This is sufficient when the only state with mass below the cutoff of our effective theory is the essentially-massless ($\mh \ll \td$) DM particle state itself.
Thus we may write
\begin{equation}\label{eq:alpha}
{\mathcal I}_k(\td/M) = A_k \left( \frac{\td}{M} \right)^{\alpha} \left(1 + \cdots \right)~,
\end{equation}
where for each process $\alpha$ is an approximately-constant exponent and $A_k$ a dimensionless pre-factor with an exact value determined by the model-dependent structure of the matrix element.  (As discussed in the conclusions the more complicated situation where ${\mathcal I}_k(\td/M)$ is not determined by a single power-law over the relevant range of $\td$ has a richer range of behaviours, as does the related case where the DM sector has massive states that can go on mass-shell
at energies ${\widetilde T}_0 > E\gg \mh$.  These interesting elaborations go beyond the scope of this work.)   For perturbative DM sectors $\alpha$ is close 
to an even integer.

As the SM temperature satisfies $\dot{T}= - H T$ ,  Eqs.~\ref{eqn:Boltzmann_n}-\ref{eq:alpha} imply that the DM yield evolves
according to
\bea\label{eq:yieldevolution}
\frac{d Y}{d T} & \simeq & - \frac{45 g (k-2)A_k}{2 \pi^2 g_*}  Z^2  \left( \frac{\Theta T}{M}\right)^{\alpha} \frac{\Theta^4}{H}\\
& \simeq & - \sqrt{\frac{256\pi^9 g_*}{ 45g^2}} \frac{ A_k (k-2) \mpl}{r^2 T^2}  \left( \frac{r T}{4 M}\right)^{\alpha} Y^{4-\alpha}~.\nonumber
\eea
Although Eq.~\ref{eq:yieldevolution} is of the form of a conventional yield evolution equation, it is less immediately useful than usually the case as the boundary conditions on the process are most naturally expressed in terms of the initial and final \emph{DM sector} temperature, which in our case evolves very differently to the SM temperature.  In particular, for purely kinematic reasons, the number-changing $2\rightarrow k$ processes certainly cease when $\td \simeq \mh$ (the number-changing mechanism can become ineffective \emph{before} this temperature if the interaction rate satisfies $\Gamma_{2\rightarrow k} < H(T)$).  

Thus it is better to track the evolution of $Y$ with respect to the DM temperature $\td$.  We find
the rate of change of the yield to be given by
\bea\label{eq:YieldTd}
\frac{dY}{d\td} &\simeq &  -\frac{Y^3}{\td} \left[Y^2 + \frac{\td}{\beta_k \mpl}\left(\frac{M}{\td}\right)^{\alpha}\right]^{-1}~,\\
{\rm with~~~}  \beta_k & \equiv & \frac{ \sqrt{80\pi^9} A_k (k-2) g_*^{1/2} }{ 15gr(1+r)^{1/2} }~.
\eea
 This equation is correct in the limit where
$Z\ll 1$ \emph{during the entire evolution} - in particular, the $k\rightarrow 2$ inverse reactions encoded by the $Z^{(k-2)}$ factor inside the brackets
in the expression Eq.~\ref{eq:collision} for the Boltzmann collision term are unimportant.  We must self-consistently check that this assumption
holds using Eq.~\ref{eq:no_cannibal} in Section \ref{sec:essential}, as we do for the phenomenological example in Section \ref{sec:pheno}.

Inspecting this equation clearly shows that the evolution of $Y$ falls into \emph{two distinct regimes}. For $Y$ sufficiently small, the yield as a function of $\td$ evolves as
\bea\label{eq:YieldTd_smallY}
\frac{dY}{d\td} &\simeq & -\beta_k \frac{\mpl}{\td^2} \left(\frac{\td}{M}\right)^{\alpha} Y^3~,
\eea
while for larger $Y$ we switch over to
\bea\label{eq:YieldTd_largeY}
\frac{dY}{d\td} &\simeq & -\frac{Y}{\td}~.
\eea
These two limiting forms of the yield equations enable a useful approximate analytic understanding of the evolution of the DM density, as we explain in the next section, while Eq.~\ref{eq:YieldTd} can be solved numerically to give an accurate value of the DM yield.

The physical interpretation of the two regimes
is as follows: the `small-$Y$' limit given by Eq.~\ref{eq:YieldTd_smallY} corresponds to the case when $Y$ develops \emph{slowly} compared to the Hubble rate, namely $\Gamma_{2\rightarrow k} \ll H$, while the `large-$Y$' scenario given by
Eq.~\ref{eq:YieldTd_largeY} corresponds to \emph{fast} evolution, $\Gamma_{2\rightarrow k} \gg H$.  
In accord with this, the solution of the fast regime evolution Eq.~\ref{eq:YieldTd_largeY} is simply 
\bea\label{eq:YieldTd_largeY_sol}
\frac{Y_f}{Y_i} &= & \frac{\td_i}{\td_f}~,
\eea
reflecting the fact that in the fast regime the red-shifting of the DM energy density via Hubble expansion is unimportant; all the available kinetic energy in the sector is efficiently processed into the mass-energy-density of the DM particles, independent of the details
of the interaction, finally ceasing when one exits the fast regime (often this is when $\td \simeq \mh$ is reached). 
On the other hand, in the slow regime, Hubble expansion plays the dominant role with both SM and DM temperatures
evolving almost perfectly in step, with the DM yield staying essentially constant as the RHS of the relevant evolution equation, Eq.~\ref{eq:YieldTd_smallY}, 
is parametrically small.   Thus, as we will see in detail, and check via a numerical solution of Eq.~\ref{eq:YieldTd}, the non-trivial evolution of the yield is essentially determined by the switch-over between the fast and slow regimes.

\section{Essential behaviours}\label{sec:essential}

The boundary between the two regimes occurs at the $\td$-dependent yield given by
\bea\label{eq:Yregimeswitch}
Y_* &=& \beta_k^{-1/2} \left(\frac{\td}{\mpl}\right)^{1/2}  \left(\frac{\td}{M}\right)^{-\alpha/2} ~.
\eea
For a given value of all the underlying interaction parameters, $A_k, \alpha$ etc, this relation defines a curve in the $\td$-$Y$ plane with parametric behaviour determined by the exponent $\alpha$.  The physics of RFI depends upon whether
the point in the plane at which particle production ceases, defined by $(\td,Y) = (m, Y_{\rm dm})$ where $Y_{\rm dm} = 0.434\eV/m$ is the yield giving the
correct DM density, is in the fast or slow regime.

\begin{figure}[h!]
  \includegraphics[scale=0.24]{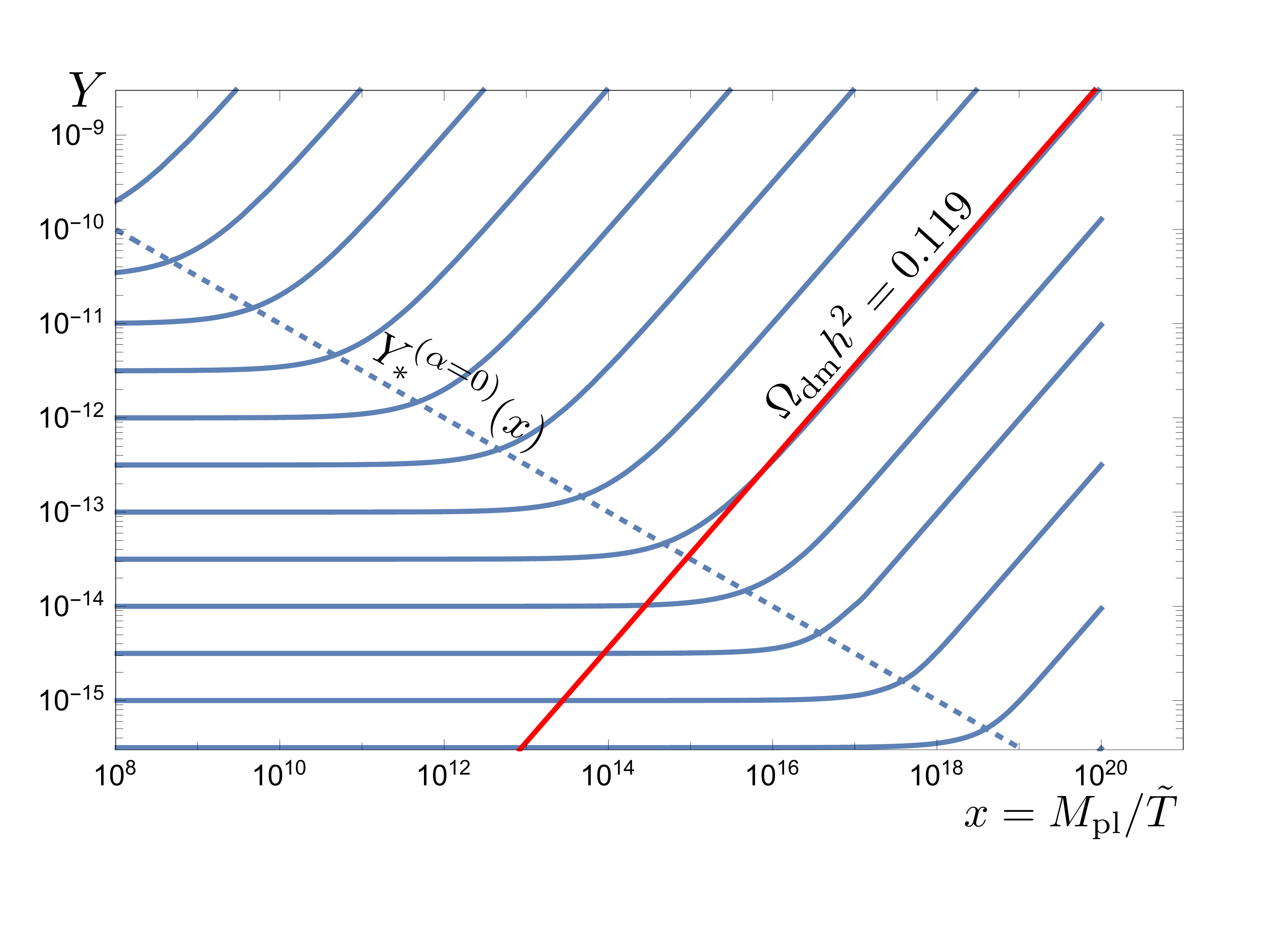}
\caption{\label{fig:Y*curve_alpha0}
	Illustration of the DM yield evolution (blue curves) as a function of $x= M_{\rm pl}/\td$ from numerical solution to Eq.~\ref{eq:YieldTd},  for the case $\alpha=0$
	and fixed interaction strength, starting with varying initial
	yields at $\td_0=10^{-8}M_{\rm pl}$.  $Y_*$ (dashed line) delineates the switch-over from slow (below) to fast (above) evolution.  
	The red line corresponds to the observed DM density.  Yield evolution terminates when $\td\simeq \mh$ (thus $x\simeq M_{\rm pl}/\mh$).  Once evolution passes into the fast regime the yield curves are iso-DM-density-lines, so on the correct yield curve {\it every} value of the DM mass gives the observed DM density (if assumptions underlying fast evolution hold). The self-consistency
	and observational constraints discussed in the text have not yet been imposed. } 
\end{figure}

\begin{figure}[h!]
  \includegraphics[scale=0.25]{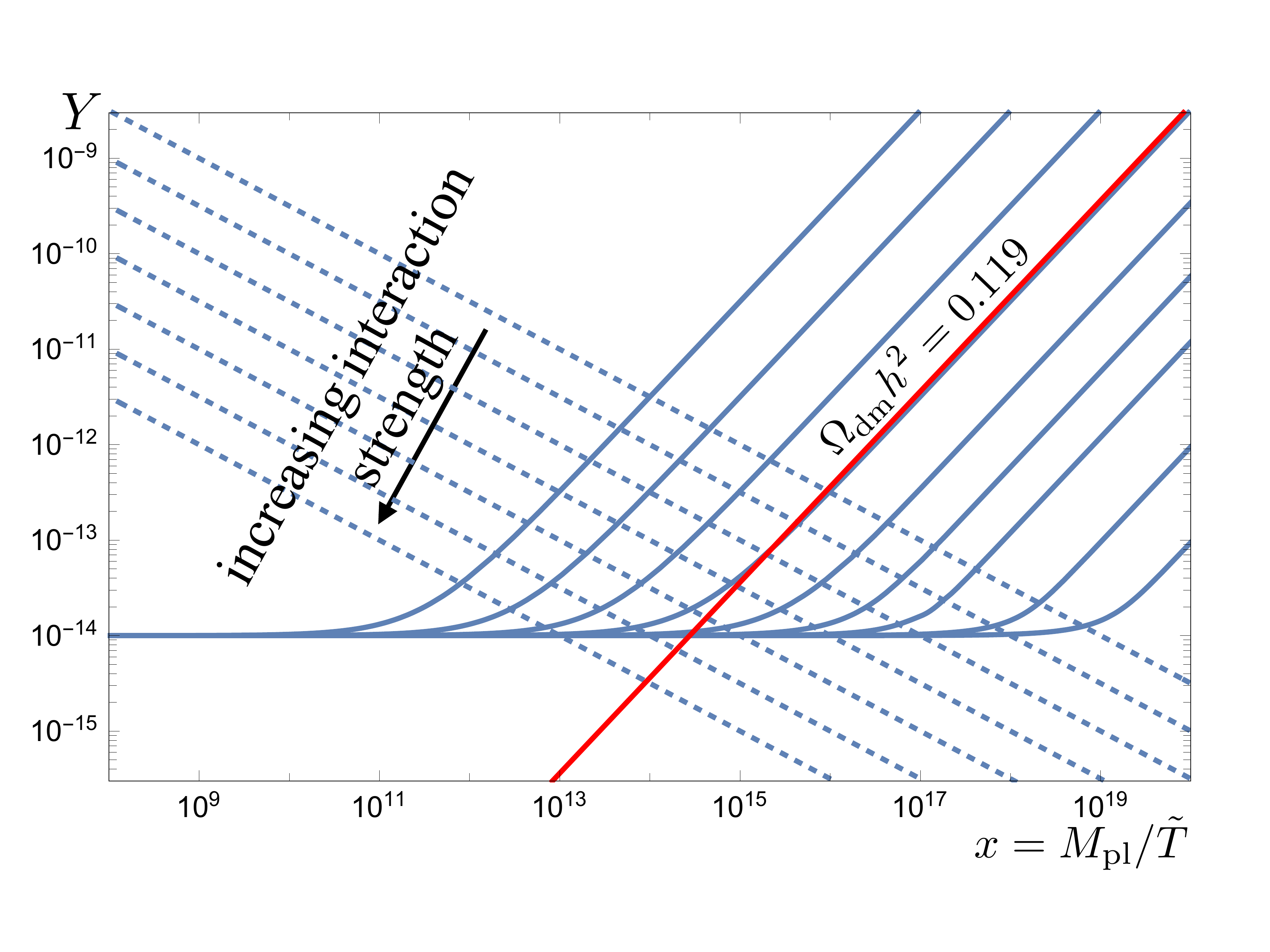}
\caption{\label{fig:Y*curve_alpha02}
	Evolution of the DM yield starting with fixed initial value $Y = 10^{-14}$ for varying interaction strengths in the case $\alpha=0$.  Conventions as in
	Fig.~\ref{fig:Y*curve_alpha0}.} 
\end{figure}

From Eq.~\ref{eq:Yregimeswitch} we see that for $\alpha < 1$ the boundary curve $Y_*(\td)$ falls as $\td$ itself decreases through
either Hubble expansion or conversion of kinetic to mass-energy.  This case corresponds to the rate of the $2\rightarrow k$ process increasing relative to the
Hubble rate as $\td$ drops.  In Fig.~\ref{fig:Y*curve_alpha0} we illustrate the behaviour of the yield as a function of
$x\equiv M_{\rm pl}/\td$ for the case $\alpha=0$ as the initial value, $Y_0$, is varied while keeping the interaction strength fixed.  In Fig.~\ref{fig:Y*curve_alpha02} we show the evolution of the yield for fixed $Y_0=10^{-14}$ in the case $\alpha=0$ as the interaction strength is increased.  In both figures the evolution should be understood to terminate
when $\td \simeq \mh$, ie, at $x_f \simeq M_{\rm pl}/\mh$.

From these figures one sees that in the case $\alpha<1$, and for sufficiently small DM mass, one always enters the fast regime where the DM yield is determined by the number-changing processes.  It is also clear that, while several curves meet the observed DM density curve (red line) in the slow region, due to the parallelism of this line to the yield flows in the fast region, only a single yield flow terminating in the fast region will produce the correct DM density. Having selected this curve, one may choose to terminate at any point after it has joined the red line, and still be confident of producing the correct relic abundance. Put another way, once $\mh < m_*$ where  $m_*$ is the value of the mass where the $ Y_*$ and $\mh Y =0.434\eV$ curves cross, DM of \emph{any} lower mass automatically gives the observed DM density if the interaction strength is correctly chosen depending on $r$ and $Y_0$.  This is the region of most interest to us.

For $\alpha=0$ and in the regime $\mh < m_*$ an approximate analytical expression relating the final yield to the underlying parameters is simply
\bea\label{eq:finalYalpha0}
Y_f &\simeq & \beta_k \frac{Y_0^3\mpl}{\mh} ~.
\eea
This shows the automatic $\mh$-independence of the final DM density $\Omega h^2 = 0.119(\mh Y_f/0.434\eV)$, but also the fact that the DM density resulting from RFI is sensitive to the initial condition, $Y_0$.  This is not such a disadvantage as it may at first seem, either because one has a predictive theory of the initial condition (as, e.g., essentially arises in primordial black hole decay\cite{Argyres:1998qn,Lennon:2017tqq,Hooper:2019gtx}), or because the idealised situation we have so far described, where the DM and SM sectors are completely decoupled from each other, is violated by a feeble coupling which is possibly measurable (e.g. via late-time decays).  Moreover,  the self-interaction that is key to the DM production is the same interaction that may in favourable cases lead to potentially observable effects in structure formation via elastic scattering.

\begin{figure}[h!]
  \includegraphics[scale=0.24]{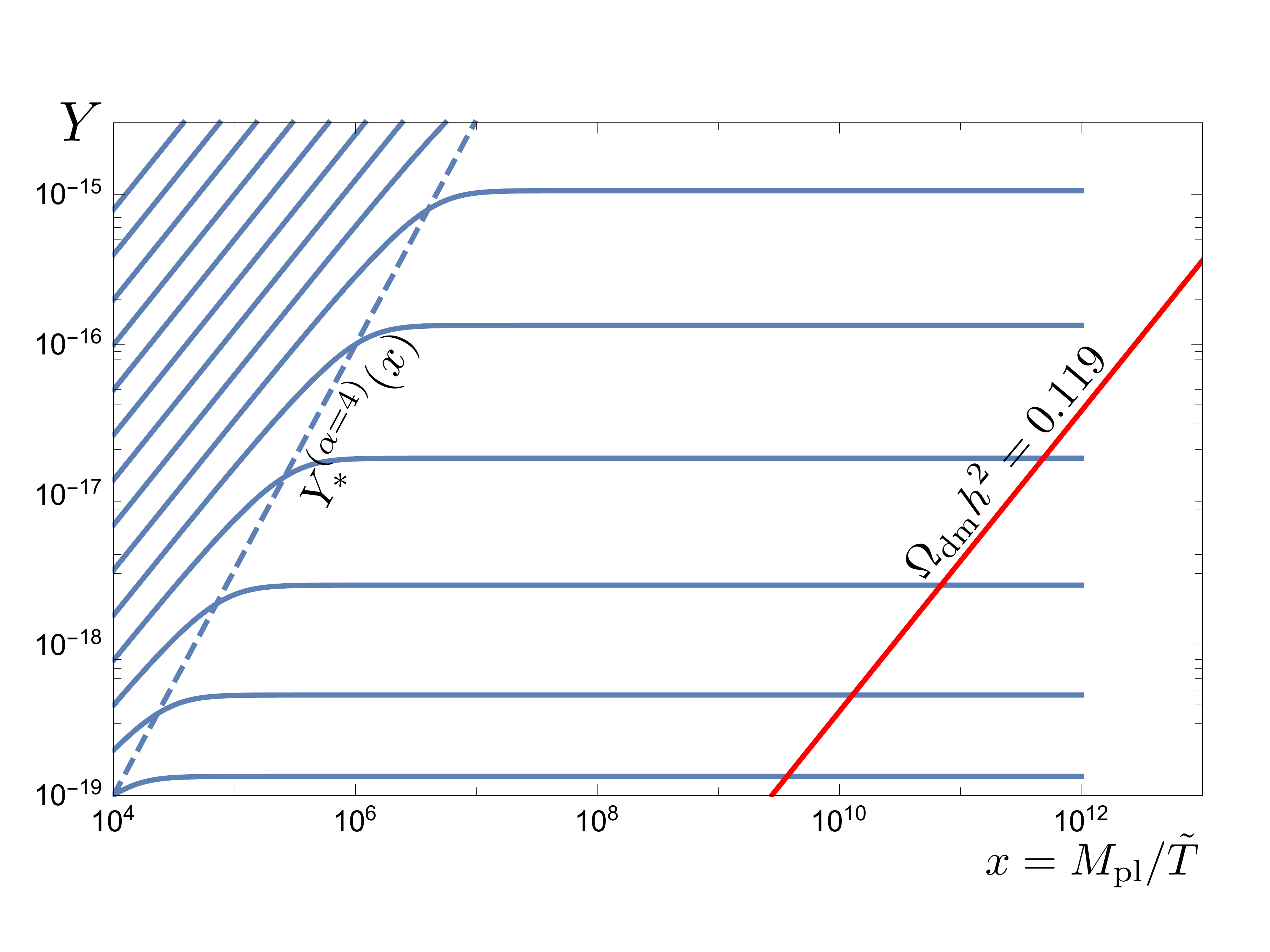}
\caption{\label{fig:Y*curve_alpha4}
	Illustration of the evolution of the DM yield for the case $\alpha=4$. Conventions as in Fig.~\ref{fig:Y*curve_alpha0}.  The self-consistency
	and observational constraints discussed in the text have not yet been imposed. 
	} 
\end{figure}

Returning to the general behaviour of the yield evolution as a function of $\alpha$,
for values of $\alpha>3$ the $Y_*$ curve increases with $x$ faster than linearly, so one finds that even if $Y_0$ is large enough so that the initial evolution
is in the fast regime where $2\rightarrow k$ processes are efficient, the evolution inevitably exits into the slow regime unless the DM mass is such that $\mh > m_*$.  This behaviour is illustrated in the case $\alpha=4$ in Fig.~\ref{fig:Y*curve_alpha4}.

\begin{figure}[h!]
  \includegraphics[scale=0.24]{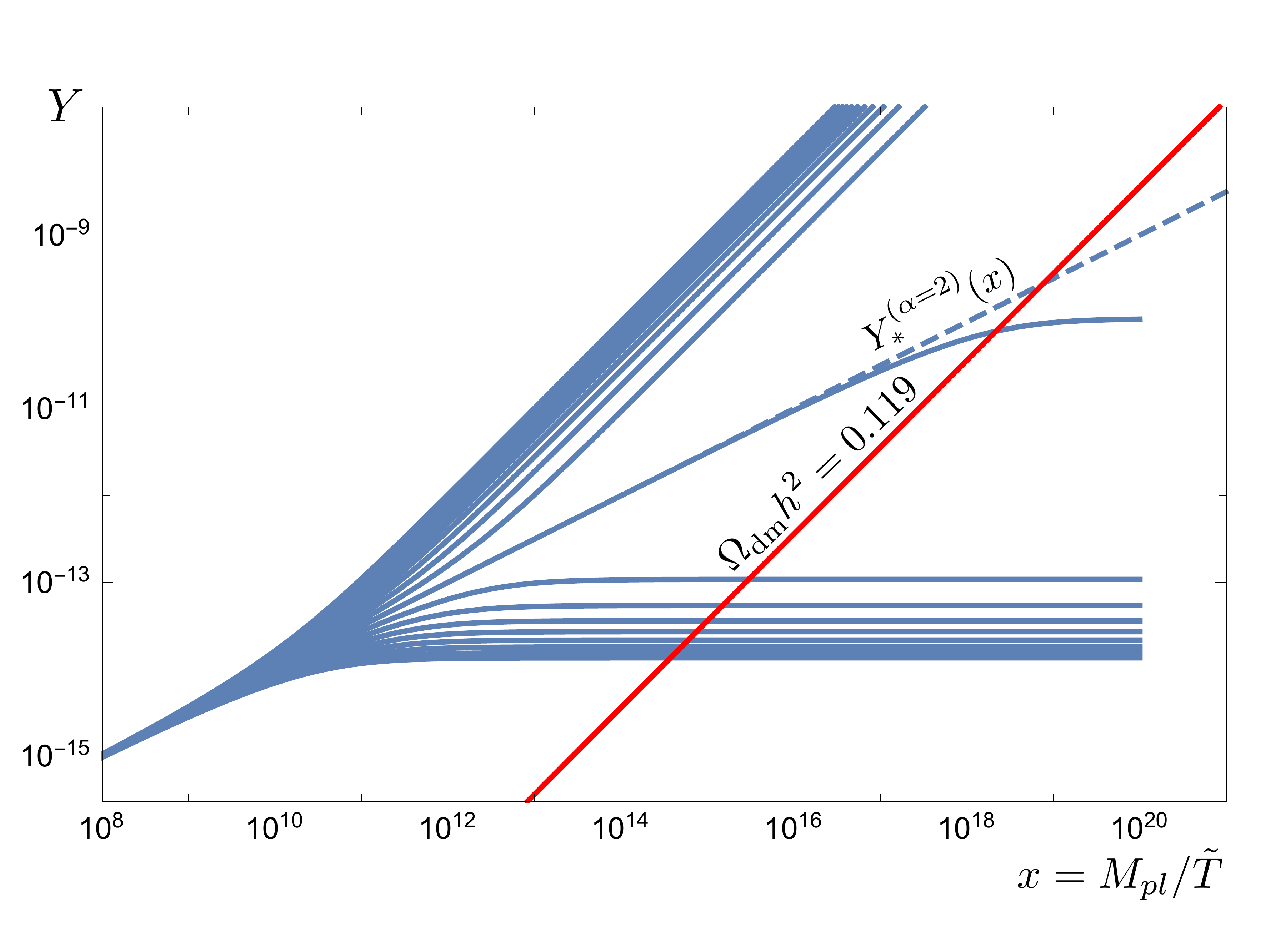}
\caption{\label{fig:Y*curve_alpha2}
	Illustration of the yield evolution for the case $\alpha=2$.   In this
	case the resulting yield curves diverge away from the $Y_*$ line marking the boundary between
	the fast and slow evolution regimes.  The self-consistency and observational
	constraints discussed in the text have not yet been imposed. 
	} 
\end{figure}

On the other hand, in the case $\alpha =2$ the yield evolution is contained fully in either the slow or fast regime (apart from a set of initial conditions
of measure zero) depending upon the initial condition for the yield, $Y_0$ and the strength of the interaction as measured by $\beta_k$.   This behaviour is illustrated
in Fig.~\ref{fig:Y*curve_alpha2}.

Finally, we must check that the neglect of the inverse $k\rightarrow 2$ processes is correct.  First this requires that the under-occupation
factor $Z\ll 1$.  Since in this limit Eqs.~\ref{eq:zvszeta} and \ref{eq:yieldtheta} imply the relation 
\bea\label{eq:Zreln}
Z_f=\frac{128 \pi^4 g_*  Y_f^4}{45g r^3} ~,
\eea
we must have the final yield value, $Y_f$, at which evolution terminates satisfy
\bea\label{eq:Yfreln}
 Y_f \ll 1\times 10^{-4} \left(\frac{r}{10^{-4}}\right)^{3/4} \left(\frac{10}{g_*/g}\right)^{1/4}~,
\eea
implying that the DM mass is bounded below by
\bea\label{eq:DMm_min}
\mh \gg 3\keV \left(\frac{r}{10^{-4}}\right)^{-3/4} \left(\frac{10}{g_*/g}\right)^{-1/4}~.
\eea
This is a sufficient condition in all cases where the yield evolution terminates in the slow regime. 

However this is not the most stringent constraint: In the cases where the evolution of the DM yield terminates (nominally, $\td \simeq \mh$)
in the regime in which the evolution is fast, then we must require that at $\td \simeq \mh$ not only is the rate for $k\rightarrow 2$ processes
much slower than the rate for $2\rightarrow k$ processes, but that it is much slower than the Hubble rate at this time.  If this is not the case
then as the DM sector temperature drops well below $\sim \mh$ the exothermic $k\rightarrow 2$ processes will still be active and fast relative to
the Hubble rate, unlike the endothermic $2\rightarrow k$ interactions which quickly become exponentially suppressed by the tiny Boltzmann
factors, and so an epoch of cannibalism of the DM number density will occur, see e.g.  \cite{Carlson:1992fn,Hochberg:2014dra,Hochberg_2015,Pappadopulo:2016pkp,Farina_2016}.  (This is, of course, not a problem for the physics - such evolution is perfectly consistent and will occur in a portion of parameter space -
rather, the DM density is now set by the freeze-out of the cannibal mechanism, with the RFI mechanism just setting the initial conditions for the later
evolution of the DM yield.)  The situation where this more stringent condition 
can be relevant is when the exponent $\alpha<3$, and is most simply expressed as
\bea\label{eq:no_cannibal}
Z_f^{(k-2)} \left(\frac{\td_*}{\mh}\right)^{(3-\alpha)} \ll 1 ~,
\eea
where $Z_f$ is given in Eq.~\ref{eq:Zreln} and $\td_*$ is the value of $\td$ where the yield evolution curves  cross the $Y_*$ line
de-marking the transition from slow to fast evolution.  If Eq.~\ref{eq:no_cannibal} is satisfied then the DM density is set by RFI and not
cannibalism.

\section{Phenomenology}\label{sec:pheno}

We now present a brief discussion of phenomenological aspects of RFI. For the sake of being definite, we focus on a simple DM sector $\lambda \phi^4 / 4!$ model with
corresponding exponent $\alpha = 0$ at leading order.   As discussed in Section~\ref{sec3}, with $\al<1$, the number-changing $2 \rightarrow k$ processes become increasingly efficient as the universe expands, terminating in the fast region when the temperature of the DM sector passes through the DM mass threshold and particle production is kinematically forbidden. The leading-order number-changing process is $2\rightarrow 4$, for which we estimate $A_4 \sim 10^{-11} \lambda^4$. 

As indicated in Section~\ref{sec:essential}, we are most interested in the case where the secluded sector evolution enters the fast regime with the DM
abundance set by Eq.~\ref{eq:finalYalpha0}. Combining this with the requirement that the final relic abundance satisfies $Y_f \mdm = 0.434$ eV, we find the useful relation
\beq\label{eq:rlamY0}
r=1.8\times10^{-8}\;\left(\frac{\lambda}{0.1}\right)^4\left(\frac{Y_0}{10^{-8}}\right)^3.
\eeq

Insisting that the sector moves into the fast regime means that there is some evolution from the initial yield $Y_0$, that is $Y_f > Y_0$. Using this condition we find an upper limit on the DM mass given by
\beq\label{eq:constr:Y0Yf}
\frac{\mh}{\GeV} \lesssim 0.0434 \left(\frac{10^{-8}}{Y_0}\right).
\eeq

Beyond the requirement that the system enters the fast regime, we also insist that the DM states do not reach
full chemical equilibrium, that is $\Theta_f > \Theta_{\rm eq}$. Combining this restriction with Eq.~\ref{eq:finalYalpha0} we find a lower limit on the mass 
\beq\label{eq:thetafgrt}
\frac{\mh}{\GeV} \gtrsim 2.0 \times 10^{-3}
\left(\frac{0.1}{\lambda}\right)^3
\left(\frac{10^{-8}}{Y_0}\right)^{9/4}~,
\eeq
where we have set $g_*=10$ for simplicity. 

The behaviour of this lower limit may initially seem unintuitive; that decreasing $\lambda$ increases the probability that equilibrium will be reached. This behaviour is, at root, due to the fact that we have imposed the requirement that the system eventually reaches the correct relic abundance: $Y_f \mh = r \mh /4\Theta_f = 0.434$ eV.  This means
that $\Theta_f \sim r$, while $\Theta_{\rm eq} \sim r^{1/4}$.  When we increase $\lambda$, because of the relationship in Eq.~\ref{eq:finalYalpha0}, this corresponds to an increase in $r$, which moves $\Theta_f$ and $\Theta_{\rm eq}$ further apart.

Recall that we must check Eq.~\ref{eq:no_cannibal} holds in order to ensure negligible rates of $4\rightarrow 2$ reactions. This is not a constraint as such - viable models outside this region are certainly possible - but it will change the prediction for the relic abundance compared with our canonical scenario due to
late-time cannibalistic reactions. To avoid this we require
\beq
\frac{\mh}{\GeV} \gg 3 \times 10^{-3} \left(\frac{0.1}{\lambda}\right)^{24/11}
\left(\frac{10^{-8}}{Y_0}\right)^{21/11}~.
\eeq

We are able to constrain the parameter space further by considering implications for structure formation, in particular by calculating the overall damping scale, $l_{\rm tot}$, up to which the matter power spectrum will be suppressed. As there are elastic scattering processes coming from the quartic interaction, there will in principle be two uncorrelated contributions to the total damping scale, one contribution from collisional damping (via diffusion) and the other from free streaming. Specifically, from the time at which the DM is first produced, $t_0$, until the $2 \rightarrow 2$ processes are overtaken by Hubble expansion, the DM will diffuse through self-interaction collisions.  After the collisions cease to be efficient, the particles stream freely until matter-radiation equality. Both of these processes lead to a smearing out of structure perturbations on scales $\lc$ and $\lFS$ respectively. 

The diffusion damping scale (squared) may be calculated using the formula in \cite{Chu:2015ipa}
\beq\label{eq:lc}
\lc^2 = \int_{t_0}^{\tdec} \frac{ \langle v \rangle^2}{a^2 n_{\rm dm} \sv}\;dt~,
\eeq
where $a$ is the scale factor, $v$ is the DM velocity, $\sigma$ is the DM self-interaction elastic scattering cross section and the angular brackets indicate thermal averaging.
The free streaming scale is \cite{Kolb:206230}
\beq\label{eq:lFSform}
\lFS = \int_{\tdec}^{t_{\rm EQ}} \frac{v}{a}\;dt~.
\eeq

In order to calculate the relevant integrands, we use the expressions previously derived for the fast regime, in the case that $\alpha = 0$. For thermally averaged quantities, we follow the procedure as set out in \cite{Edsj__1997}.  We find that, for our regions of interest, the collisional damping scale is dominated by diffusion that takes place while the population is non-relativistic.

If the DM self-interaction decoupling takes place before the time of matter-radiation equality at $\TSM = \Teq$, we have the following expression for $\lc$:
\beq
\begin{aligned}
	 \lc & \simeq1.0\times 10^{-3} \kpc
	 \left(\frac{\lambda}{0.1}\right)^4 \left(\frac{Y_0}{10^{-8}}\right)^3
	 %\\
	 %&\simeq 5.5 \kpc\;\; \left(\frac{r}{10^{-4}}\right)
	 ~,
\end{aligned}
\eeq
and free streaming scale is calculated to be
\beq
\begin{aligned}
	\lFS &\simeq 4.4 \times 10^{-4} \kpc\; \left(\frac{\lambda}{0.1}\right)^4\left(\frac{Y_0}{10^{-8}}\right)^3\\
	&\times
	\left(
        28 + \log \left[\left(\frac{\mh}{\GeV}\right)^3 
	\left(\frac{0.1}{\lambda}\right)^6 
	\left(\frac{10^{-8}}{Y_0}\right)^3
 	\right]
	\right)
%	\\
%        &\simeq 2.4\times \kpc \left(\frac{r}{10^{-4}}\right) \\
%        &  \times \left(
%	19 + \log \left[ \left(\frac{\mh}{\GeV}\right)^3 \left(\frac{10^{-4}}{r}\right) \left(\frac{0.1}{\lambda}\right)^2\right]
%	\right)
	~.
\end{aligned}
\eeq

The scales are uncorrelated and therefore are to be combined in quadrature \cite{Chu_2015} and the total required to be less than $\sim 100$ kpc in comoving units to compare favourably with numerical structure formation simulations \cite{Boehm:2004th,Boehm} i.e.:
\beq\label{eq:ltot}
l_{\rm tot} = \sqrt{l_c^2 + \lFS^2} \lesssim 100 \text{ kpc}.
\eeq

If elastic scattering decouples \textit{after} matter-radiation equality, collisional damping runs down to $\Teq$ and free-streaming never contributes, giving the total damping length as
\beq
\begin{aligned}
l_{\rm tot} \simeq 2 \times 10^{-3}\kpc \left( \frac{\mh}{\GeV}\right)^{3/2} \left( \frac{Y_0}{10^{-8}}\right)^{3/2} \left( \frac{\lambda}{0.1}\right) 
%\\
%\simeq 8.8 \times 10^{-3}\kpc \left(\frac{\mh}{\GeV}\right)^{3/2} 
%\left(\frac{r}{10^{-4}}\right)^{1/2} \left( \frac{0.1}{\lambda}\right)
~.
\end{aligned}
\eeq

It is worthy of note that we find it is only for a subset of $Y_0$ values that the total length scale can ever exceed $100$ kpc.  That is to say, outside of this range, there is no choice of $\mh$ and $\lambda$ consistent with the other constraints which gives an $\ltot > 100$ kpc.  We thus need only concern ourselves with the damping
scales for $Y_0 \in (\sim 7 \times 10^{-10},  \sim 10^{-7})$. This will be evident from the plots we discuss below. 

We point out that we may also expect  oscillatory signatures in the matter power spectrum \cite{Das:2018ons}, although such an analysis is beyond the scope of this work.

Finally, a constraint may be derived from the bound on DM self-interactions from cluster collisions, e.g. the Bullet Cluster and other collisions, and the observed properties of the DM halos of galaxies, see e.g. \cite{Markevitch_2004, Randall:2007ph, Kahlhoefer:2013dca,Tulin:2017ara}.  The Bullet cluster itself imposes a limit of $\sigma_{\rm elas}/\mh \lsim 1.25{\rm cm}^2/{\rm g}$, while mildly stronger constraints $\sigma_{\rm elas}/\mh \lsim 0.5{\rm cm}^2/{\rm g}$ arise from the other observations, including halo ellipticity, although there is in principle a considerable uncertainty in these limits \cite{Vogelsberger:2012ku,Rocha:2012jg,Peter:2012jh,Zavala:2012us,Tulin:2017ara}, with some claims that there exists positive evidence for DM self-interactions.
To be conservative we impose the stronger bound which translates into the constraint
\bea\label{eq:clusters}
\frac{\mdm}{\GeV}\gsim 1.0 \times 10^{-2} \lambda^{2/3}~.
\eea
It is worth noting that this constraint is independent of $Y_0$ and, given Eq.~\ref{eq:constr:Y0Yf}, as $Y_0$ is decreased the impact of the self-interaction limit on the allowed parameter space will become increasingly marginal as the upper limit on the DM mass increases. 

\begin{figure}[t!]
	\includegraphics[width=.5\textwidth]{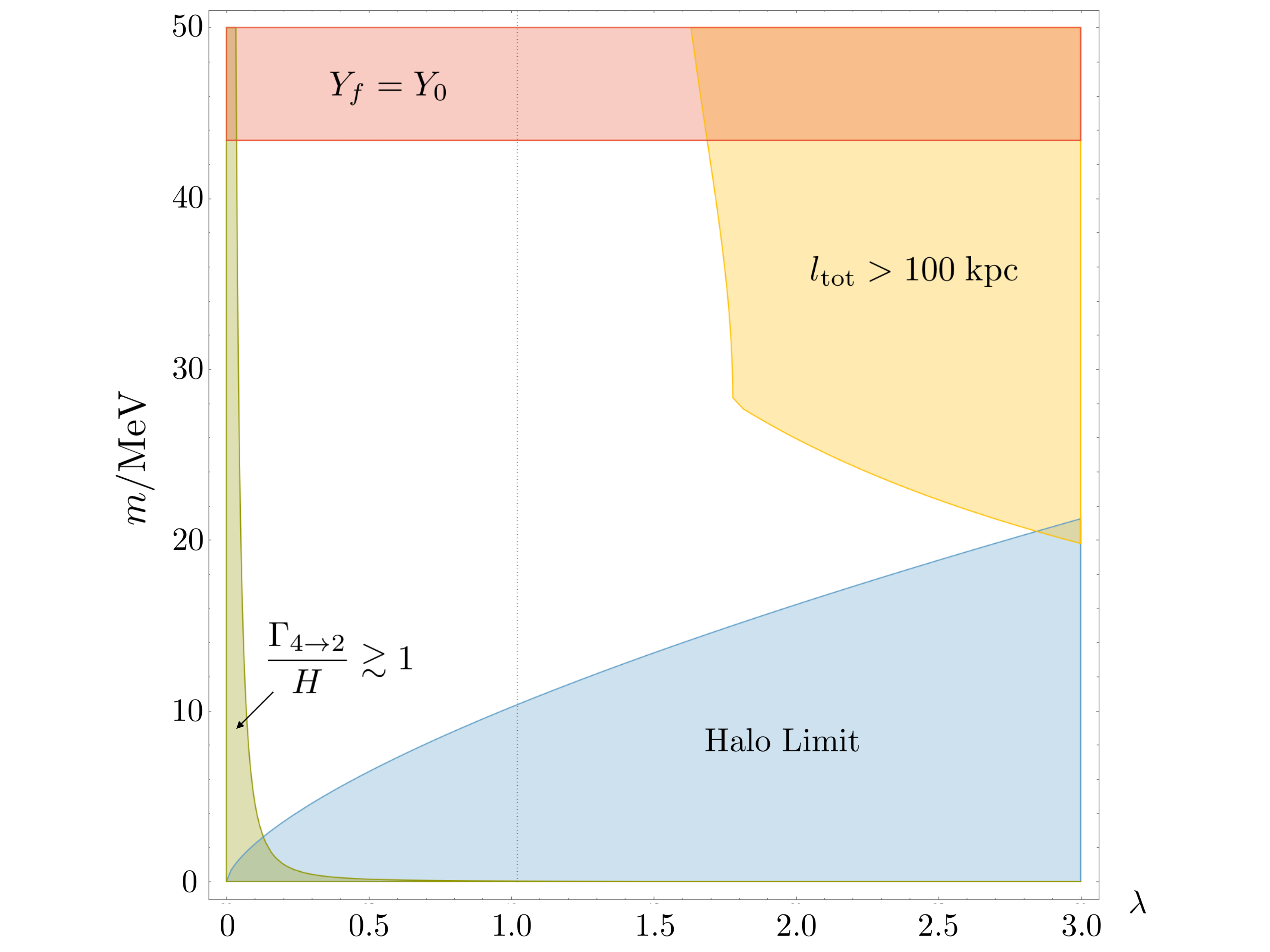}
	\caption{The $\lambda-\mh$ parameter space for $Y_0 = 10^{-8}$. The red region eliminates those points for which $Y_0 = Y_f$, the blue those which violate the bound in Eq.~\ref{eq:clusters}, the orange those for which $\ltot > 100$ kpc, the dark green those for which cannibalistic processes may occur.  The region to the right of the light vertical dotted line would be excluded in the case of no self-interactions, see text for details.}
	\label{fig:mlamplot}
\end{figure}

\begin{figure}[t!]
	\includegraphics[width=.5\textwidth]{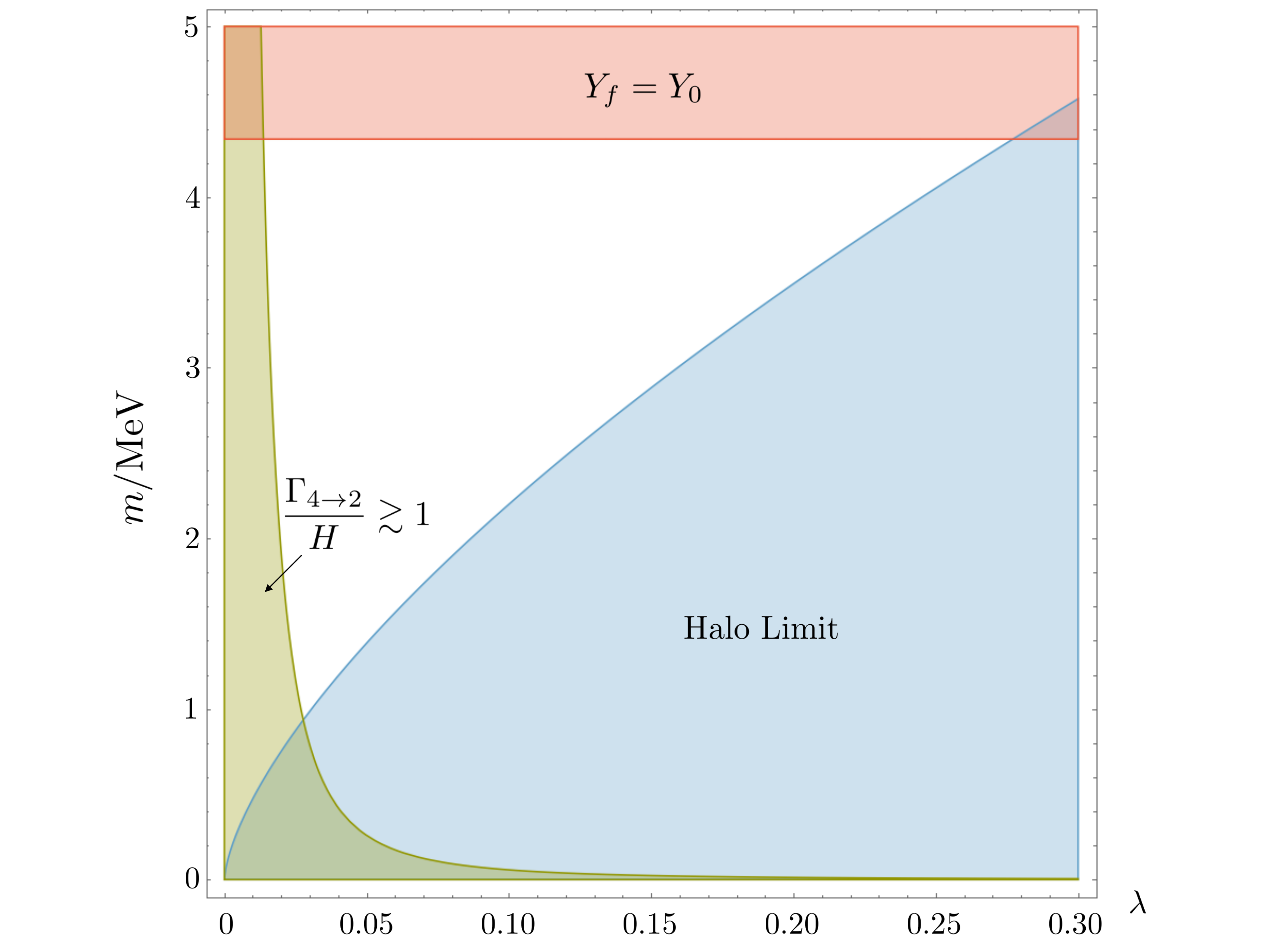}
	\caption{The $\lambda-\mh$ parameter space for $Y_0 = 10^{-7}$. Conventions as described in Fig.~\ref{fig:mlamplot}.}
	\label{fig:mlamplot2}
\end{figure}

\begin{figure}[h]
	\includegraphics[width=.5\textwidth]{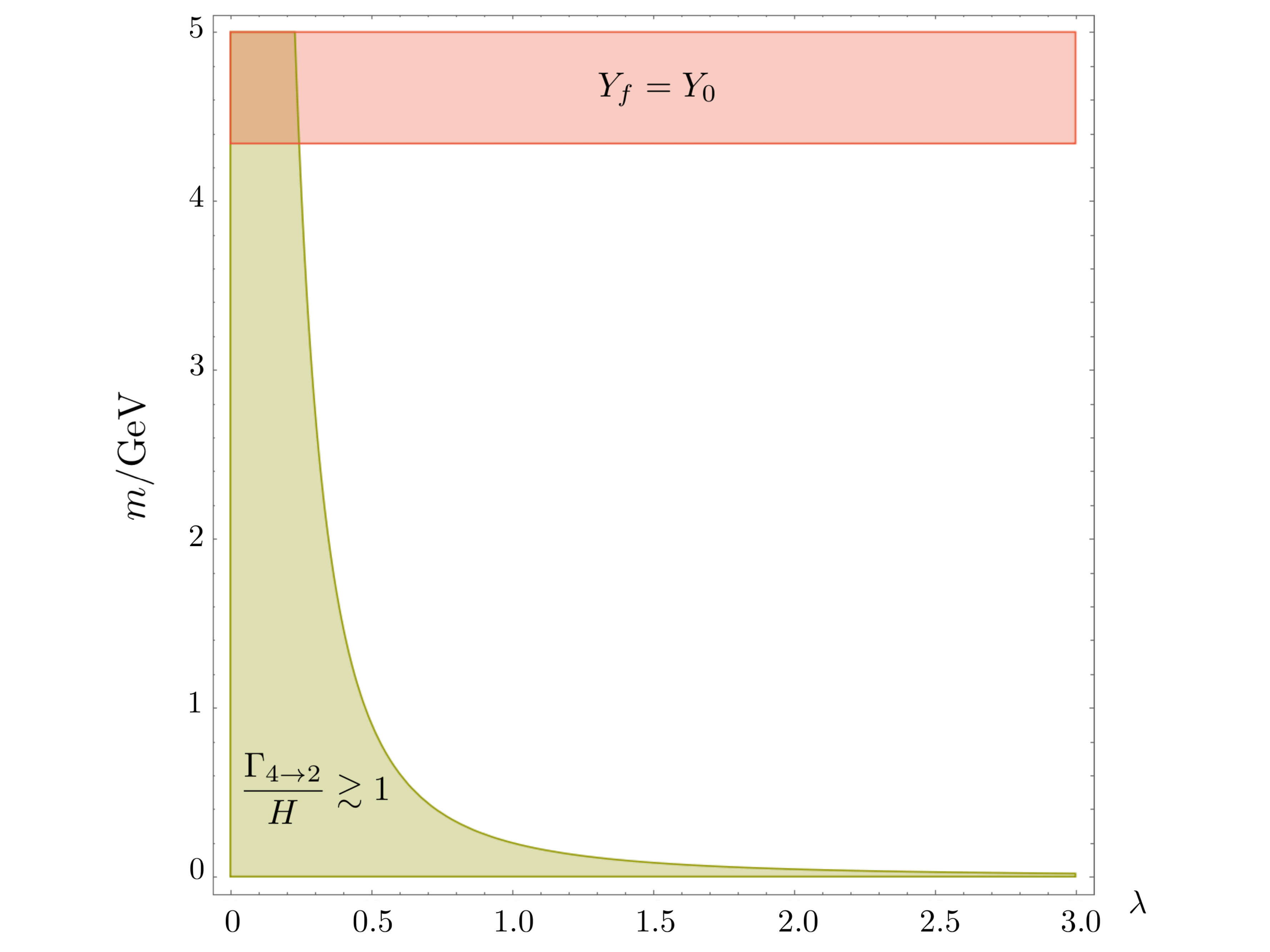}
	\caption{The $\lambda-\mh$ parameter space for $Y_0 = 10^{-10}$.  Conventions as described in Fig.~\ref{fig:mlamplot}.}
	\label{fig:mlamplot3}
\end{figure}

In Figs.~\ref{fig:mlamplot}, \ref{fig:mlamplot2}, and \ref{fig:mlamplot3}  we summarise the constraints on the $\lambda-\mh$ plane for this particular model of RFI with $\alpha=0$ for different choices of the initial value of the yield, $Y_0=10^{-8}, 10^{-7}, 10^{-12}$ respectively. 

In Fig.~\ref{fig:mlamplot}, with $Y_0=10^{-8}$ the $\lambda$-independent upper bound on the mass (red region) coming from the requirement that some evolution in the fast region occurs is given by Eq.~\ref{eq:constr:Y0Yf} to be $\mh<43.4\;$MeV.  The exclusion regions corresponding to the self-interaction ``Halo Limit" (blue area) and restrictions on the damping scales (orange area) rule out the large $\lambda$ values apart from a wedge around $\mh\sim 20\;$MeV.  The requirement that the number-changing $4\rightarrow 2$ interactions do not become important (green area) rules out a slice at small $\lambda$. 

The sharp corner to the damping-excluded orange-coloured region indicates the crossover at which the elastic-scattering decoupling temperature falls below $\Teq$, which to a first approximation switches the damping off at a relatively earlier scale.  Also indicated in Fig.~\ref{fig:mlamplot} is a dashed vertical line.  This line shows where the damping scale limit would fall, everything to the right being ruled out, if there were no self-interactions, that is, if the DM was able to free stream from early times.

Although in the plot we show the $\lambda-\mh$ plane we can convert easily to limits on $r$ for a given $Y_0$ by using Eq.~\ref{eq:finalYalpha0}. For example, from Fig.~\ref{fig:mlamplot} we can deduce that the maximum value of $r$ permitted for $Y_0=10^{-8}$ is $r\sim 2 \times 10^{-2}$.

In Fig.~\ref{fig:mlamplot2}, the $\lambda-\mh$ parameter space for $Y_0 = 10^{-7}$ is displayed. As discussed above, the limits from structure formation no longer impinge on the parameter space.  This is due to the factor of ten reduction in the upper limit on the DM mass now appearing at $\mh = 4.3\;$MeV and the limits from the halo limit in this mass range being much stronger than those coming from the damping scale.  The maximum value for $r$ for $Y_0=10^{-7}$ is $r\sim 10^{-3}$.

In Fig.~\ref{fig:mlamplot3} we show the $\lambda-\mh$ parameter space for $Y_0 = 10^{-10}$.  With $Y_0$ being small the upper limit on the DM
mass increases to $\mh = 4.3\;$GeV.  The constraints from structure formation are not relevant in this case due to the small $Y_0$ and the halo limit is only relevant for a very tiny slice at low masses not visible in the plot.  There are two limits on the maximum value of $\lambda$, one coming from BBN (combining Eqs.~\ref{eq:constr:DNeff} and \ref{eq:rlamY0}) and the second from the perturbativity of $\lambda$.  The perturbativity limit (not displayed in Fig.~\ref{fig:mlamplot3}) is the strongest of these two with the maximum value for $r$ in this case being $r\sim 5\times10^{-6}$.

\section{Conclusions and remarks}\label{sec:conclusions}

In this work we have presented the RFI mechanism for DM production. DM resides in an initially ultra-relativistic, far under-populated secluded sector and undergoes a period of rapid number density increase via $2\to k$ processes.  This converts the initially large DM kinetic energy to mass, in the form of additional DM states. The relic abundance is set by the point at which the DM temperature drops below its mass at which stage the $2\to k$ ($k>2$) processes become inefficient. 

We have made some simplifying assumptions that have allowed us to present a straightforward analytic understanding of the mechanism. Most fundamentally, we have so far assumed that \emph{kinetic} equilibrium is quickly established in the secluded sector by elastic interactions, so that the distribution functions are determined solely by an evolving temperature and chemical potential.  This enabled us to write down a simple evolution equation for the DM yield, Eq.~\ref{eq:YieldTd}.  However, the RFI mechanism does \emph{not} require that full kinetic equilibrium is established.  The distribution functions could have significant non-thermal `tails' or other features which
change the resulting DM yield, or alter the astrophysical and cosmological signatures in interesting ways (for instance there could remain a small ultra-relativistic sub-component
of the DM until late times), though such an investigation goes beyond the scope
of this work.  Even if our simplifying assumption of kinetic equilibrium is a good approximation, there are a number of issues that deserve investigation. 
Important aspects to follow up on include the thermal corrections to the DM masses and couplings.  Although these considerations will not modify the general picture of the mechanism, they may change some of the detailed behaviour, in particular where the evolution tracks the ultra-relativistic secluded sector.   Late-time exothermic `cannibal' interactions are also a natural possibility in this class of models.
In this treatment we have assumed that $\gst$ and thus $r$ are constant all the way down to $\Teq$. The formalism of RFI freely
admits a more detailed approach in which $\gst$ and thus $r$ vary over time. 

Again for simplicity we have assumed that there is one dominant number changing interaction, with one fixed temperature dependence, that is, one particular value of $\alpha$. Often this will not be the case.  For example, if the secluded sector consists of multiple particles the temperature dependence of the collision term, Eq.~\ref{eq:alpha}, may change dramatically as the DM temperature drops below mass thresholds of states involved in the number changing interactions driving RFI. Resonance effects may also
play a material role.

Since this work was concerned with exploring the general features of the RFI mechanism, we haven't studied specific implementations
beyond the $\lambda \phi^4$ model examined in Section~\ref{sec:pheno}.  However we emphasise that many models previously considered
in the context of self-interacting or cannibal DM, such as hidden-sector `pion' DM \cite{Bai:2010qg,Bhattacharya:2013kma,Kribs:2016cew} 
and possibly glueball DM \cite{Faraggi:2000pv,Boddy:2014yra,Garcia:2015loa,Soni:2016gzf,Kribs:2016cew},
also have a regime of parameter space where the RFI mechanism determines
the DM density.  The confinement transition, though, must be carefully examined in the non-equilibrium case of a
far-under-populated, high-temperature sector, a topic which we hope to return to in a later work.

The DM self-interactions studied in this paper could also play a role in alleviating the present $\sigma_8$ and Hubble tensions \cite{Bernal:2016gxb,Aghanim:2018eyx,Riess:2019cxk}, due to such
effects as a DM viscosity \cite{Anand:2017wsj}, or due to the collisional-damping/free-streaming studied in Section~\ref{sec:pheno}, and/or a possible
sub-component of the DM (which could either be the non-thermalised `tail' mentioned above, or a new light state mediating the elastic and inelastic
interactions) acting as dark radiation (see e.g., \cite{Dessert:2018khu}).    

Finally we note that in this paper we have strictly forbidden any SM-DM interactions.  As we discussed, the mechanism is still testable by virtue of the fact that the same self-interaction is responsible for the DM number changing-interactions that determine the relic abundance and the elastic scattering processes that may be potentially observable in, or at least constrained by, structure formation. If, however, we relax this restriction on SM-DM interaction it may be possible for this mechanism to admit decaying DM signals \cite{Mardon:2009gw} (note the DM can be superheavy $\mh \gg \TeV$ as long as it was also initially ultra-relativistic), or other interesting effects.

\section*{Acknowledgments}

JMR gratefully thanks the Galileo Galilei Institute for Theoretical Physics and the Simons Foundation for hospitality and support during this work, and Daniel Egana-Ugrinovic for discussions.  HT thanks the Science and Technology Facilities Council (STFC) for a postgraduate studentship. SMW thanks the University of Oxford Physics Department for hospitality during the completion of this work and is funded by STFC Grant No. ST/P000789/1.

%\bibliography{rfi.bib}
\input{rfi.bbl}

\end{document}